\newcommand{\bea}{\begin{eqnarray}}
\newcommand{\eea}{\end{eqnarray}}
\newcommand{\be}{\begin{equation}}
\newcommand{\ee}{\end{equation}}
\newcommand{\ben}{\begin{enumerate}}
\newcommand{\een}{\end{enumerate}}
\newcommand{\bi}{\begin{itemize}}
\newcommand{\ei}{\end{itemize}}
\newcommand{\bmi}[1]{\begin{minipage}{#1 cm}}
\newcommand{\emi}{\end{minipage}}

\newcommand{\rund}[1]{\left(#1\right)}
\newcommand{\vc}[1]{\mbox{\boldmath $#1$}}
\renewcommand{\d}{{\rm d}}
\newcommand{\eck}[1]{\left[ #1 \right]}
\newcommand{\ave}[1]{\left\langle #1 \right\rangle}
\newcommand{\vt}{\vartheta}
\newcommand{\vp}{\varphi}
\newcommand{\abs}[1]{| #1 |}

\def\llabel#1{\label{sc:#1}}
\def\elabel#1{\label{eq:#1}}

{\catcode`\@=11
\gdef\SchlangeUnter#1#2{\lower2pt\vbox{\baselineskip 0pt \lineskip0pt
  \ialign{$\m@th#1\hfil##\hfil$\crcr#2\crcr\sim\crcr}}}
}
\def\gtrsim{\mathrel{\mathpalette\SchlangeUnter>}}
\def\lesssim{\mathrel{\mathpalette\SchlangeUnter<}}

\documentclass[onecolumn]{aa}
\usepackage{graphicx}
\usepackage{epsfig}
\begin{document}
   \title{Galaxy-galaxy-galaxy lensing: Third-order correlations between
      the galaxy and mass distributions in the Universe}

   \author{
  Peter Schneider
          \and
          Peter Watts  
          }

   \offprints{P. Schneider}

   \institute{Institut f. Astrophysik u. Extr. Forschung, Universit\"at Bonn,
              Auf dem H\"ugel 71, D-53121 Bonn, Germany\\
              \email{peter@astro.uni-bonn.de, pwatts@astro.uni-bonn.de}
             }
\titlerunning{Galaxy-galaxy-galaxy lensing}

   \date{Received ; accepted }

   \abstract{Galaxy-galaxy lensing (GGL) measures the 2-point cross-correlation between
   galaxies and mass in the Universe. In this work we seek to generalise this
   effect by considering the {\em third-order}  correlations between galaxies and mass:
   galaxy-galaxy-galaxy lensing. Third-order correlations in the cosmic shear
   field have recently been reported in the VIRMOS-DESCART and CTIO surveys. 
   Such data should also be 
   ideal for measuring galaxy-galaxy-galaxy lensing. Indeed, the
   effects of these higher-order correlations may have already been detected
   in recent studies of galaxy-galaxy lensing. Higher-order cross-correlation
   functions contain invaluable information about the relationship between
   galaxies and their mass environments that GGL studies alone cannot detect. 

   In this paper we lay out the basic relations for third-order cross
   correlations and their projections and introduce a new set of scale
   dependent third-order bias parameters. We define three new observables: two
   galaxy-shear-shear correlation functions, $G_{\pm}$, and a
   galaxy-galaxy-shear correlation, ${\cal G}$. We relate these to the
   various projected cross-bispectra and give practical estimators for their
   measurement. We note that the observational signature of these
   correlators is simply the excess shear-shear correlation measured about
   foreground galaxies (for $G_{\pm}$) and the average tangential shear around
   foreground galaxy pairs (for $\cal G$). These quantities are no more than
   second order in the shear and so should be more easily measurable than the
   shear 3-point correlation. Finally we derive expressions for the third
   order aperture mass statistics in terms of both the cross-bispectra and
   the real-space correlation functions. Such statistics provide a very
   localized measurement of the bispectra, thus encapsulating
   essentially all of the available third-order information, while remaining
   easily obtainable from observations of 3-point cross-correlation
   functions. In addition we find that utilising aperture statistics has the
   further benefit that they measure only the {\em connected} part of the third
   order correlation. 

   }

   \maketitle
%

\section{\llabel{intro}Introduction}

Gravitational lensing offers the possibility of studying the statistical
properties of the mass distribution in the Universe without referring
to the relation between the mass and the luminous tracers, like
galaxies. The weak lensing effect (see Mellier 1999; Bartelmann \&
Schneider 2001, hereafter BS01; Refregier 2003; van Waerbeke \&
Mellier 2003 for recent reviews)
of the large-scale structure, called
cosmic shear (see, e.g., Blandford et al.\ 1991; Miralda-Escud\'e
1991; Kaiser 1992; Jain \& Seljak 1997), has been detected by a number
of groups (e.g. van
Waerbeke et al.\ 2000, 2001, 2002; Bacon et al.\ 2000; Kaiser et al.\ 2000;
Wittman et al.\ 2000; Maoli et al.\ 2001; Refregier et al.\ 2002; H\"ammerle
et al.\ 2002; Brown et al.\ 2003; Jarvis
et al.\ 2003). Most of these cosmic shear measurements have concentrated on
second-order shear statistics, but quite recently, third-order cosmic shear
measurements have been reported as well (Bernardeau et al.\ 2002;
Pen et al.\ 2003; Jarvis et al.\ 2004).

Weak lensing can also be used for measuring the relation between mass
and galaxies. The measurement of this galaxy-galaxy lensing (GGL) was first
attempted by Tyson et al.\ (1984), and its first detection was
published by Brainerd et al.\ (1996). Since then, quite a number of
GGL measurements have been reported, in the HST Medium
Deep Survey (Griffiths et al.\ 1996), the Hubble Deep Field
(Dell'Antonio \& Tyson 1998; Hudson et al.\ 1998), and more recently,
in the RCS survey by Hoekstra et al.\ (2001, 2002), the COMBO-17 survey
(Kleinheinrich et al.\ 2004) and the Sloan Digital Sky Survey (Fischer
et al.\ 2000; McKay et al.\ 2001; Guzik \& Seljak 2002; Sheldon et
al.\ 2004, Seljak et al.\ 2004). GGL measures the two-point
correlation function (2PCF) between galaxy positions and shear.

On small scales, GGL measures the lensing effect of the dark halo in
which the luminous galaxy is embedded. It therefore provides a probe of
the density profile of galaxy halos which can be compared to the
predictions of the CDM model. In particular, the properties of dark
matter halos, such as virial mass and concentration, can be studied as
a function of galaxy type (early vs. late), environment (e.g., have
the halos of galaxies in groups and clusters been stripped by the
tidal interactions), and redshift. On scales larger than $\sim 200
h^{-1}\,{\rm kpc}$, the GGL signal is no longer dominated by the halo
of galaxies, but receives a substantial contribution from the matter in
which galaxies are embedded, namely groups and clusters. From this
signal, one learns about the environments of galaxies. For example,
the density-morphology relation can be traced back directly to the
underlying mass density, instead to the galaxy number density.

Schneider (1998) pointed out that the correlation between galaxy
positions and shear provides a direct measure of the bias factor of
galaxies; in particular, the scale dependence of the bias factor can
be probed directly (van Waerbeke 1998). Indeed, on scales $\gtrsim 2
h^{-1}\,{\rm Mpc}$, GGL measures a combination of the galaxy-mass
correlation coefficient $r$ and the bias factor $b$. These two functions
can be obtained individually when the GGL signal is combined with the
cosmic shear signal, as shown by Hoekstra et al.\ (2002). In fact,
Hoekstra et al.\ (2002) have measured $r$
and $b$ as a function of scale. Future measurements of
these quantities as a function of scale, redshift and galaxy type,
using the upcoming wide-field imaging surveys, will provide invaluable
input for the interpretation of galaxy redshift samples. In addition,
$b$ and $r$ are functions which depend on the formation and evolution
of galaxies; they can therefore be used as constraints for those
models.

As already mentioned, a non-vanishing cosmic shear three-point
correlation function (3PCF) has been detected, and future
surveys will measure it with great precision. These surveys will
certainly also measure higher-order correlations between galaxies
and matter, generalizing the GGL measurements. Although the
detailed physical interpretation of such higher-order cross
correlations remains to be investigated, it is obvious that they
contain valuable 
information concerning the relationship between galaxies and their
mass environments {\em beyond} that which is measurable in GGL. In
the language of the halo occupation distribution (Berlind \&
Weinberg 2002), one can think of these higher-order functions as
probing the moments of an occupation probability $P(N|M)$ that a
halo of virial mass $M$ contains $N$ galaxies of a particular
type. Such information is of crucial importance since it places
powerful constraints upon models of galaxy formation.

We save the detailed interpretation of higher-order cross
correlations for a forthcoming paper, where we explore such themes
in the context of the halo model. In this work we lay out the
basic relations for the third-order galaxy-mass correlations.
After introducing the projected mass and galaxy number densities
in Sect.\ts 2, we define the third-order bias factor and two
third-order correlation coefficients in Sect.\ts 3. The relation
between the projected bispectra and the three-dimensional
bispectra is given in Sect.\ts 4.  Practical estimators for the
galaxy-galaxy-shear and the galaxy-shear-shear correlation
functions are provided in Sect.\ts 5. We will argue that these
third-order galaxy-mass correlations have already been measured
and published. In Sect.\ts 6, the relation between the correlation
functions and the cross-bispectra are derived, and a consistency
relation for the two galaxy-shear-shear 3PCFs is obtained that should
be satisfied provided the shear is a pure E-mode field. 
In Sect.\ts 7 we define aperture measures of the
projected matter and galaxy densities and consider their
third-order statistics; in particular, on the one
hand they are related to the projected bispectra, on the other hand,
they can be 
directly calculated from the respective correlation functions. We
discuss our results in Sect.\ts 8.

M\'enard et al.\ (2003) have considered a different approach to measure
third-order correlations between the mass and the galaxy distribution,
employing the magnification bias of background galaxies. Generalizing an
effect that has been observed at the two-point statistical level, where
correlations between high-redshift QSOs and low redshift galaxies are observed
and which are interpreted as being due to the magnification caused by the
large-scale matter distribution of which the foreground galaxies are biased
tracers (see Dolag \& Bartelmann, 1997; Bartelmann \& Schneider 2001, and
references therein), M\'enard et
al.\ (2003) considered the excess of distant QSOs around pairs of foreground
galaxies. For linear deterministic bias of the galaxy distribution, the ratio
of this third-order statistics and the square of the QSO-galaxy correlation
function becomes essentially independent of the bias, as well as of the shape
and amplitude of the power spectrum of density fluctuations. This result is
very similar to the corresponding ratio of third-order cosmic shear statistics
and the square of the shear dispersion (Bernardeau et al.\ 1997; Schneider et
al.\ 1998; van Waerbeke et al.\ 1999).

\section{\llabel{proje}Projection of density and galaxies}
The gravitational lensing effect of the inhomogeneous matter distribution in
the Universe is described in terms of an equivalent surface mass density
$\kappa(\vc\theta)$, which is obtained by projecting the three-dimensional
density contrast, $\delta$, of the matter along the line-of-sight (see
BS01, and references therein). If we
consider sources at comoving distance $w$, this surface mass density is given
by
\be
\kappa(\vc\theta,w)=\frac{3H_0^2\Omega_{\rm m}}{2c^2}\,
  \int_0^w\,\d w'\,\frac{f_K(w')f_K(w-w')}{f_K(w)}\,
  \frac{\delta\rund{f_K(w')\vc\theta,w'}}{a(w')}\;,
\elabel{6.21}
\end{equation}
where $f_K(w)$ is the comoving angular-diameter distance corresponding
to comoving distance
$w$, $f_K(w)=w$ for flat Universes, $\delta=\Delta\rho/\bar\rho$ is the
relative density contrast of matter, $H_0$ is the Hubble constant, $c$ the
velocity of light, $a(w)=1/(1+z)$ is the cosmic scale factor, normalized to
unity today, and $\Omega_{\rm m}$ is the density parameter in matter.
For a redshift distribution of (`background') sources with probability density
$p_z(z)\,\d z=p_w(w)\,\d w$,
the effective surface mass density becomes
\be
\kappa(\vc\theta)=\int\d w\;p_w(w)\,\kappa(\vc\theta,w)
=\frac{3H_0^2\Omega_{\rm m}}{2c^2}\,
  \int_0^{w_{\rm h}}\d w\;g(w)\,f_K(w)\,
\frac{\delta\rund{f_K(w)\vc\theta,w}}{a(w)} \;,
\elabel{5.11}
\ee
with
\be
g(w)=\int_w^{w_{\rm h}}\d w'\;p_w(w'){f_K(w'-w)\over f_K(w')}\;,
\elabel{g-fact}
\ee
which is the source-redshift weighted lens efficiency factor
$D_{\rm ds}/D_{\rm s}$ for a
density fluctuation at distance $w$. The quantity
$w_{\rm h}$ is the comoving horizon distance, obtained from $w(a)$ by letting
$a\to 0$. The shear, i.e. the projected tidal gravitational field which can be
measured as the expectation value of image ellipticities of the galaxy
population, is obtained from the projected surface mass density $\kappa$ in
the usual way: one defines the deflection potential $\psi(\vc\theta)$ which
satisfies the two-dimensional Poisson equation $\nabla^2\psi=2\kappa$, in
terms of which the Cartesian components of the shear read
$\gamma_1=(\psi_{,11}-\psi_{,22})/2$, $\gamma_2=\psi_{,12}$, where indices
separated by a comma denote partial derivatives.

In analogy to the matter density contrast $\delta$, we define the number
density contrast
$\delta_{\rm g}(\vc x,w)$ of galaxies as
\be
 \delta_{\rm g}(\vc x,w):={n(\vc x,w)-\bar n(w) \over \bar n(w)}\;,
\elabel{deltagal}
\ee
where $n(\vc x,w)$ is the number density of galaxies at comoving position $\vc
x$ and comoving distance $w$ (the latter providing a parameterization of cosmic
time), and $\bar n(w)$ is the mean number density of galaxies at that
epoch. Since the galaxy distribution is discrete, the true number density is
simply a sum of delta-functions. What is meant by $n$ is that the probability
of finding a galaxy in the volume $\d V$ situated
at position $\vc x$ is $n(\vc x)\,\d V$.
We consider now a population
of (`foreground') galaxies with spatial number density $n(\vc x,w)$.
The number density of these galaxies on the sky at $\vc\theta$ is then
$N(\vc\theta)=\int\d w\; \nu(w)\, n(f_k(w)\vc\theta,w)$, where $\nu(w)$ is the
redshift-dependent selection function, describing what fraction of the
galaxies at comoving distance $w$ is included in the sample.
The selection function $\nu(w)$ can differ between different types of
foreground galaxies and thus depends on the sample selection.
The mean number
density of these galaxies on the sky is
$\bar N=\int\d w\;\nu(w)\,\bar n(w)$; the redshift distribution of these
galaxies, or more precisely, their distribution in comoving distance therefore
is $p_{\rm f}(w)=\nu(w)\,\bar n(w)/\bar N$, thus relating the
selection function $\nu(w)$ to the redshift distribution. Using the
definition (\ref{eq:deltagal}), one then finds for the number density of
galaxies $N(\vc\theta)$ on the sky and their fractional number density contrast
$\kappa_{\rm g}(\vc\theta)$,
\be
N(\vc\theta)=\bar N\eck{1+\int\d w\;p_{\rm f}(w)\,\delta_{\rm
    g}(f_K(w)\vc\theta, w)}\; ; \;\;
\kappa_{\rm
  g}(\vc\theta):=\eck{N(\vc\theta) - \bar N}/\bar N = \int\d w\;p_{\rm
f}(w)\,\delta_{\rm g}(f_K(w)\vc\theta, w) \;.
\elabel{galdensi}
\ee

\section{\llabel{bias}Definition of bias and correlation coefficients}
\subsection{\llabel{SOS}Second-order statistics}
In its simplest form, the relation between the density contrast
$\delta$ and the fractional number density contrast $\delta_{\rm
g}$ of galaxies is described by the bias factor $b$, in the form
$\delta_{\rm g}=b\,\delta$. This picture of linear deterministic
biasing is most likely too simple a description for the galaxy
distribution, at least on small spatial scales where the density
field is non-linear. Instead, one defines the bias parameter in
terms of the power spectra of the galaxy and matter distribution.
Let $\hat \delta(\vc k)$ and $\hat\delta_{\rm g}(\vc k)$ be the
Fourier transforms of the density and galaxy contrast,
respectively; their power spectra $P_{\delta \delta}$ and $P_{\rm
g g}$ are defined by the correlators
\be
\ave{\hat\delta(\vc
k,w)\,\hat\delta^*(\vc k',w)}=(2\pi)^3 \delta_{\rm D}(\vc k-\vc
k')\,P_{\delta \delta}(|\vc k|,w) \; ;\;\; \ave{\hat\delta_{\rm
g}(\vc k,w)\,\hat\delta_{\rm g}^*(\vc k',w)}=(2\pi)^3 \delta_{\rm
D}(\vc k-\vc k')\,P_{\rm g g}(|\vc k|,w) \; ,
\ee
where $\delta_{\rm D}$ denotes Dirac's delta-`function', and we have
allowed for an evolution of the power spectra with redshift or
comoving distance $w$. The occurrence of the delta function is due
to the assumed statistical homogeneity of the random matter and
galaxy fields, and the fact that the power spectra depends only on
the modulus of $\vc k$ is due to their statistical isotropy. The
bias parameter $b(k,w)$ is defined by the ratio of these two power
spectra,
    \be
        P_{\rm g g}(k,w)=b^2(k,w)\,P_{\delta \delta}(k,w)\;,
        \elabel{bias-def}
    \ee
which agrees with the previous definition in the case of linear
deterministic biasing, but is far more general. In particular, we
allow for a scale- and redshift dependence of $b$. Next, we define
the cross-power spectrum $P_{\delta{\rm
    g}}(k,w)$ through
\be \ave{\hat\delta(\vc k,w)\,\hat\delta_{\rm g}^*(\vc
k',w)}=(2\pi)^3 \delta_{\rm D}(\vc k-\vc k')\,P_{\delta{\rm
g}}(|\vc k|,w) \; ;. \elabel{crossPdef} \ee The cross-power spectrum is related to
the power spectrum of the matter density by
    \be
        P_{\delta\rm g}(k,w)=b( k,w)\,r(k,w)\,P_{\delta \delta}(k,w)\;,
        \elabel{Pdeltag}
    \ee
where we have defined the galaxy-mass correlation coefficient
$r(k,w)$, which can also depend on scale and redshift. In the case
of linear deterministic biasing, $r\equiv 1$.

\subsection{\llabel{TOS}Third-order statistics}
We now generalize the foregoing definitions to third-order statistics. The
bispectra of the matter and galaxy distributions
are defined through the
triple correlators
    \bea
        \ave{\hat\delta(\vc k_1)\,\hat\delta(\vc k_2)\,\hat\delta(\vc k_3)}
        &=&(2\pi)^3\,\delta_{\rm D}(\vc k_1+\vc k_2+\vc k_3)\,
        B_{\delta\delta\delta}(\vc k_1,\vc k_2,\vc k_3;w)\; , \nonumber \\
        \ave{\hat\delta_{\rm g}(\vc k_1)\,\hat\delta_{\rm g}(\vc k_2)\,
        \hat\delta_{\rm g}(\vc k_3)} &=&(2\pi)^3\,\delta_{\rm D}(\vc
        k_1+\vc k_2+\vc k_3)\, B_{\rm ggg}(\vc k_1,\vc k_2,\vc k_3;w)\; ,
    \eea
where the delta function ensures that these triple correlators
vanish unless the three $\vc k$-vectors form a closed triangle --
this property again follows from the statistical homogeneity of
the density fields. Furthermore, statistical isotropy causes the
bispectrum to depend only on the length of two of its $\vc
k$-vectors and the angle they enclose. Finally, parity invariance
of the density fields requires that the bispectra are even
functions of this angle. In a similar spirit to Eq.\thinspace
(\ref{eq:crossPdef}) we can define the cross-bispectra 
    \bea
        \ave{\hat\delta(\vc k_1)\,\hat\delta(\vc k_2)\,\hat\delta_{\rm
        g}(\vc k_3)} &=&(2\pi)^3\,\delta_{\rm D}(\vc k_1+\vc k_2+\vc
        k_3)\, B_{\delta\delta\rm g}(\vc k_1,\vc k_2;\vc k_3;w)\; , \nonumber \\
        \ave{\hat\delta_{\rm g}(\vc k_1)\,\hat\delta_{\rm g}(\vc k_2)\,
        \hat\delta(\vc k_3)} &=&(2\pi)^3\,\delta_{\rm D}(\vc k_1+\vc k_2+\vc k_3)\,
        B_{{\rm gg}\delta}(\vc k_1,\vc k_2;\vc k_3;w)\; .
    \eea
where we have introduced the notation $\rund{\vc k_1,\vc k_2;\vc
k_3}$ to indicate the symmetry with respect to
interchanging the first two arguments, meaning, e.g.,  that 
$B_{{\rm gg}\delta}(\vc k_1,\vc k_2;-\vc k_1-\vc k_2)
=B_{{\rm gg}\delta}(\vc k_2,\vc k_1;-\vc k_1-\vc k_2)$.

In the case of linear deterministic biasing, the various bispectra
would be related through simple relations, e.g., $B_{\delta\rm
gg}=b^2\,B_{\delta\delta\delta}$. However, in the more general
(and realistic) case, these relations are more complex. We define
the third-order bias parameter $b_3$ and the two galaxy-mass
correlation coefficients $r_1$ and $r_2$ through the relations
    \bea
        B_{\rm ggg}(\vc k_1,\vc k_2,\vc k_3;w) &=&
        b_3^3(\vc k_1,\vc k_2,\vc k_3;w)\,
        B_{\delta\delta\delta}(\vc k_1,\vc k_2,\vc k_3;w)\;, \nonumber \\
        B_{{\rm gg}\delta }(\vc k_1,\vc k_2;\vc k_3;w)
        &=& b_3^2(\vc k_1,\vc k_2,\vc k_3;w)\, r_2(\vc k_1,\vc k_2;\vc
        k_3;w)\, B_{\delta\delta\delta}(\vc k_1,\vc k_2,\vc k_3;w)\;, \\
        B_{\delta\delta\rm g}(\vc k_1,\vc k_2;\vc k_3;w)
        &=& b_3(\vc k_1,\vc k_2,\vc k_3;w)\, r_1(\vc k_1,\vc k_2;\vc k_3;w)\,
        B_{\delta\delta\delta}(\vc k_1,\vc k_2,\vc k_3;w)\;.
    \eea
The bias and correlation coefficients satisfy the same symmetries as
those of the bispectra. In the case of linear deterministic biasing,
$b_3=b$ and $r_1=1=r_2$. These higher-order biasing and correlation
parameters depend on the formation and evolution of galaxies and their
relation to the underlying density field; in general, they encapsulate
information concerning the non-Gaussian nature of the bias. As we show
in the next section, the correlation of lensing shear with the
positions of galaxies on the sky provides a direct means for measuring
(integrals of) these higher-order functions.

\section{\llabel{projected} Projected power- and bispectra}
The lensing effect and the position of galaxies on the sky are both described
by projections of the density field $\delta$ and the galaxy field $\delta_{\rm
  g}$ along the line-of-sight, as shown in Sect.\ts\ref{sc:proje}. The
projected fields share the properties of the three-dimensional distributions
as being homogeneous and isotropic random fields; hence, their second- and
third-order properties can as well be described by power and bispectra.
The relation between the power spectra of projected quantities to that of the
three-dimensional distribution is given by Limber's equation in Fourier space,
as has been derived by Kaiser (1992): Let
\be
\kappa_i(\vc\theta)=\int\d w\;q_i(w)\,\delta_i\rund{f_K(w)\vc\theta,w}
\elabel{T3-15}
\ee
be projections of the 3-D fields $\delta_i$, where the $q_i(w)$ are weight
functions depending on the comoving distance. Provided these weight functions
do not vary appreciably over comoving scales on which the power spectrum is
significantly different from zero, the (cross) power spectrum reads
\be
P_{12}(\ell)=\int\d w \;{q_1(w)\, q_2(w)\over f_K^2(w)}\,
P_\delta\rund{{\ell\over f_K(w)},w}  \;.
\elabel{limber}
\ee
Hence, the 2-D power at
angular scale $2\pi/\ell$ is obtained from the 3-D power at
length scale $f_K(w)\,(2\pi/\ell)$, integrated over $w$.
For the power spectra of the projected mass density $\kappa$ and the galaxies
$\kappa_{\rm g}$, and the cross-power of these two quantities, we therefore
find, by comparing Eqs.\ (\ref{eq:T3-15}) and (\ref{eq:limber}) with
(\ref{eq:5.11}) and (\ref{eq:galdensi}),
    \bea
        P_{\kappa \kappa}(\ell) &=& \frac{9H_0^4\Omega_{\rm m}^2}{4c^4}\,
        \int_0^{w_{\rm h}}\d w\,\frac{g^2(w)}{a^2(w)}\,
        P_{\delta \delta}\rund{k;w}\; ; \;\;
        P_{\rm g g}(\ell) = \int \d w\;{p_{\rm f}^2(w)\over
        f_K^2(w)}\,b^2\rund{k;w}\,P_{\delta \delta} \rund{k;w} \;;
        \nonumber \\
        P_{\kappa{\rm g}}(\ell) &=& {3 H_0^2\,\Omega_{\rm m}\over 2 c^2}
        \int\d w\;{g(w)\,p_{\rm f}(w)\over f_K(w)\,a(w)} b\rund{k;w}
        r\rund{k;w} P_{\delta \delta}\rund{k;w}
        \;, \elabel{6.25}
    \eea
where $\vc k=\vc \ell/f_K(w)$ and where we have used the
definition of the bias and galaxy-mass parameters given in
Sect.\ts\ref{sc:SOS}. The quantity $P_{\kappa \kappa}(\ell)$ is
the power spectrum probed by second-order cosmic shear
measurements whereas $P_{\rm g g}(\ell)$ is the power spectrum of
the galaxy distribution [with the selection function specified by
$p_{\rm f}(w)$] on the sky, i.e., the Fourier transform of the
angular 2PCF of galaxies. The cross-power $P_{\kappa\rm g}(\ell)$
is the quantity that is measured in galaxy-galaxy lensing
observations; see Hoekstra et al.\ (2002).

The bispectrum of the projected quantities (\ref{eq:T3-15}) is defined in
terms of the triple correlator
\be
\ave{\hat\kappa_1(\vc\ell_1)\hat\kappa_2(\vc\ell_2)\hat\kappa_3(\vc\ell_3)} =
(2\pi)^2\,\delta_{\rm D}(\vc\ell_1+\vc\ell_2+\vc\ell_3)\,
b_{123}(\vc\ell_1,\vc\ell_2,\vc\ell_3)\;, \elabel{angbispec}
\ee
where the relation between the bispectrum of the projected quantities $\kappa$
and the 3-D bispectrum is given by
\be
b_{123}(\vc\ell_1,\vc\ell_2,\vc\ell_3)
=\int \d w\;{q_1(w) q_2(w) q_3(w)\over f_K^4(w)}
\,B_{123}\rund{{\vc\ell_1\over f_K(w)},{\vc\ell_2\over f_K(w)},{\vc\ell_3\over
    f_K(w)};w} \;,
\ee
which is valid under the same conditions required for the validity of
(\ref{eq:limber}), i.e., the weight functions $q_i(w)$ should not vary
appreciably over scales on which the bispectrum is significantly
non-zero. Inserting into this relation the weight functions $q$ appropriate
for the surface mass density $\kappa$ and the projected number density
$\kappa_{\rm g}$, one obtains the bispectra
    \bea
        b_{\kappa\kappa\kappa}(\vc\ell_1,\vc\ell_2,\vc\ell_3)
        &=& {27 H_0^6 \Omega_{\rm m}^3\over 8 c^6}
        \int \d w\;{g^3(w)\over f_K(w) a^3(w)}\,
        B_{\delta\delta\delta}(\vc k_1,\vc
        k_2,\vc k_3;w)\;,  \\
        b_{\kappa\kappa\rm g}(\vc\ell_1,\vc\ell_2;\vc\ell_3)
        &=& {9 H_0^4 \Omega_{\rm m}^2\over 4 c^4}
        \int \d w\;{g^2(w)\, p_{\rm f}(w)\over f_K^2(w) a^2(w)}\, \, \big[b_3\,r_1
        \, B_{\delta\delta\delta}\big](\vc k_1,\vc k_2;\vc k_3;w)
        =:\eck{\bar b_3 \, \bar r_1 \, 
b_{\kappa\kappa\kappa}}(\vc\ell_1,\vc\ell_2;\vc\ell_3)
        \;,  \\
        b_{\rm {gg} \kappa}(\vc\ell_1,\vc\ell_2;\vc\ell_3)
        &=& {3 H_0^2 \Omega_{\rm m}\over 2 c^2}
        \int \d w\;{g(w)\, p_{\rm f}^2(w)\over f_K^3(w) a(w)}\, \, \big[b_3^2 \, r_2
        \, B_{\delta\delta\delta}\big](\vc k_1,\vc k_2;\vc k_3;w)\;
        =:\eck{\bar b_3^2 \, \bar r_2 \, 
b_{\kappa\kappa\kappa}}(\vc\ell_1,\vc\ell_2;\vc\ell_3)
        \\
        b_{ \rm ggg}(\vc\ell_1,\vc\ell_2,\vc\ell_3) &=& \int \d w\;{p_{\rm
        f}^3(w)\over f_K^4(w)}\, \, \big[b_3^3 \, B_{\delta\delta\delta}\big](\vc
        k_1,\vc k_2,\vc k_3;w) =: \eck{\bar b_3^3
        \, b_{\kappa\kappa\kappa}}(\vc\ell_1,\vc\ell_2,\vc\ell_3) \;,
    \eea
where again the vectors $\vc k_i=\vc \ell_i/f_K(w)$, and we have
defined the projected bias factor $\bar b_3$ and correlation
coefficients $\bar r_1$, $\bar r_2$, which depend, in addition to the
angular wave vectors $\vc\ell_i$, on the redshift distribution of
foreground and background galaxies. The correlations are strongest
if the redshift distribution of the foreground galaxies matches
the most efficient lensing redshift for a given source redshift
distribution -- see Hoekstra et al.\ (2002) for a similar
discussion in the context of second-order galaxy-mass
correlations.

\section{Estimators}
The basic observables for the second-order statistics obtained
from observational data are the 2PCFs. Hence, the angular 2PCF of
galaxies $\omega(\theta)=\ave{\kappa_{\rm
    g}(\vc\phi) \kappa_{\rm g}(\vc\phi+\vc\theta)}$ is the Fourier transform
of $P_{\rm g g}(\ell)$. The second-order statistics of the
projected matter density are probed by the shear 2PCFs
$\xi_\pm(\theta)$. They are defined in the following way: let
$\gamma_{\rm c}=\gamma_1+{\rm i}\gamma_2$ denote the complex shear
in Cartesian coordinates. Given a direction $\vp$, we define the
rotated shear $\gamma(\vc \theta;\vp)=-{\rm e}^{-2{\rm
i}\vp}\gamma_{\rm
  c}(\vc\theta)$. This definition implies that
$\gamma(\vc\theta;\vp_1) =\gamma(\vc\theta;\vp_2)\,{\rm e}^{2{\rm
i}(\vp_2-\vp_1)}$. The real and imaginary part of
$\gamma(\vc\theta;\vp)$ are the tangential and cross component of
the shear relative to the direction $\vp$ (see, e.g., Crittenden
et al.\ 2002; Schneider et al.\ 2002). Hence the shear 2PCFs
are written as
    \be
        \xi_+(\theta)=\ave{\gamma(\vc\phi;\vp)
        \gamma^*(\vc\phi+\vc\theta;\vp)} \; ;\;\;
        \xi_-(\theta)=\ave{\gamma(\vc\phi;\vp)
        \gamma(\vc\phi+\vc\theta;\vp)} \;, \elabel{XIpm}
    \ee
where the average is over all pairs of points with angular
separation $\theta$, and $\vp$ is the polar angle of their
separation vector $\vc\theta$. The imaginary parts of $\xi_\pm$
vanish due to parity invariance (e.g., Schneider 2003). For a set
of observed galaxies, the shear in (\ref{eq:XIpm}) is replaced by
the image ellipticities, which are an unbiased estimator of the
shear (we neglect here the difference between shear and reduced
shear; see BS01). The shear 2PCFs are related in a simple way to
the power spectrum $P_{\kappa \kappa}(\ell)$ -- see Kaiser (1992).

The cross-power is probed by the galaxy-galaxy lensing signal, which
correlates the tangential component of the shear at the location of a
background galaxy with the position of a foreground galaxy. We define this
correlation function as
\be
\ave{\gamma_{\rm t}}(\theta)=\ave{\kappa_{\rm
    g}(\vc\phi)\,\gamma(\vc\phi+\vc\theta;\vp)}
\elabel{GGL}
\ee
(e.g., Hoekstra et al.\ 2002),
where $\vp$ is, as before, the polar angle of the connection vector
$\vc\theta$, and the imaginary part of (\ref{eq:GGL}) vanishes due to parity
invariance. A practical estimator for $\ave{\gamma_{\rm t}}$ is obtained by
averaging the tangential shear component of the background galaxies in all
foreground-background pairs with separation $\theta$.

Similarly, the third-order statistics are best probed from
observing the 3PCFs. The angular 3PCF of galaxies is one of the
standard ways to characterize the non-Gaussian properties of the
galaxy distribution on the sky, and standard methods for
estimating it are known (see, e.g., Peebles 1980). The bispectrum
$b_{\kappa\kappa\kappa}$ is 
probed by the shear 3PCF, as defined in Schneider \& Lombardi
(2003; see also Takada \& Jain 2003a; Zaldarriaga \& Scoccimarro
2003), and its relation to the bispectrum has been derived in
Schneider et al.\ (2004). We next derive expressions for the
correlation functions related to the cross-bispectra.

\begin{figure}
\vspace{0.5cm}
   \centering
   \epsfig{file=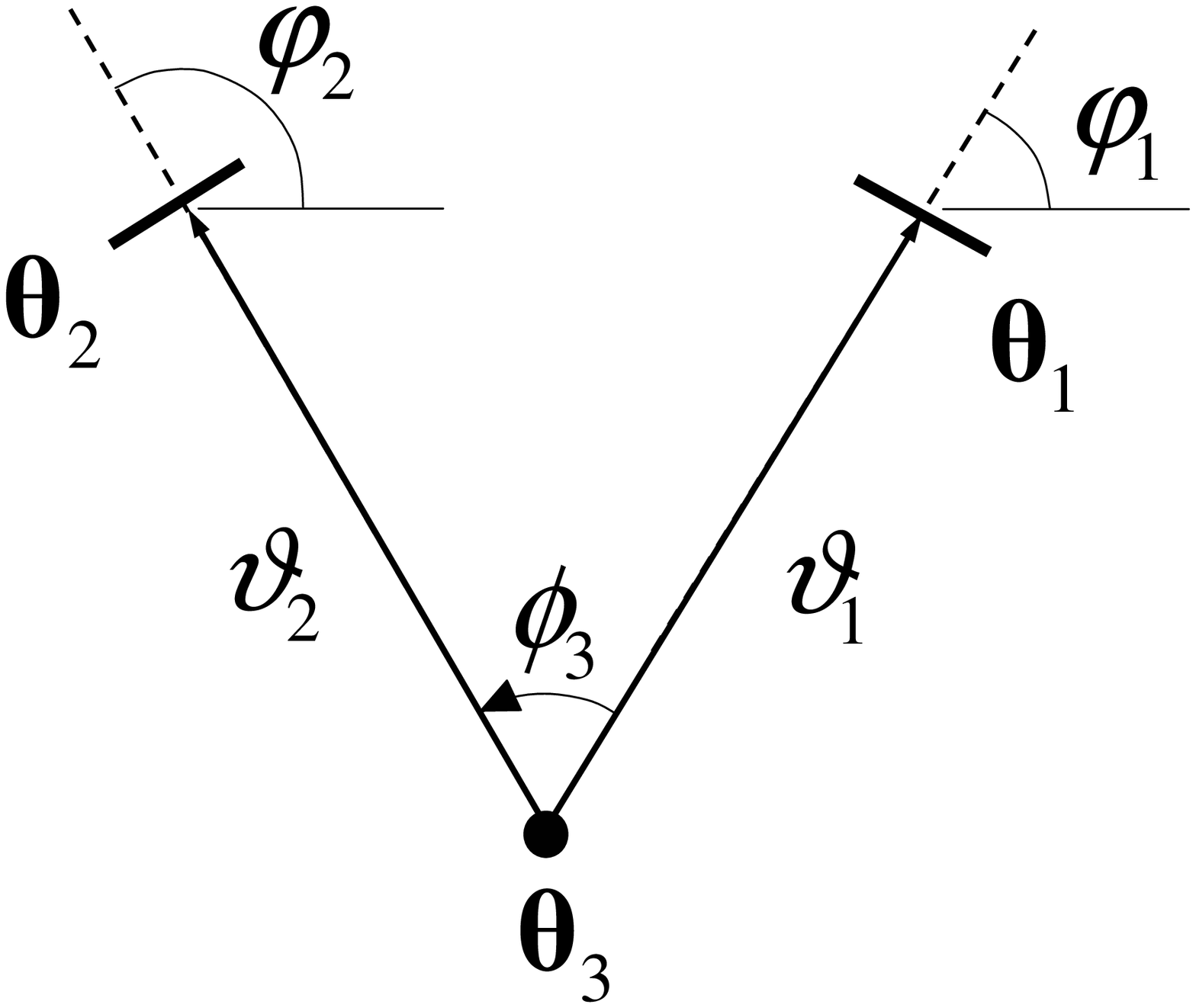,width=6.8cm,clip=}
   \hspace{2.5cm}
   \epsfig{file=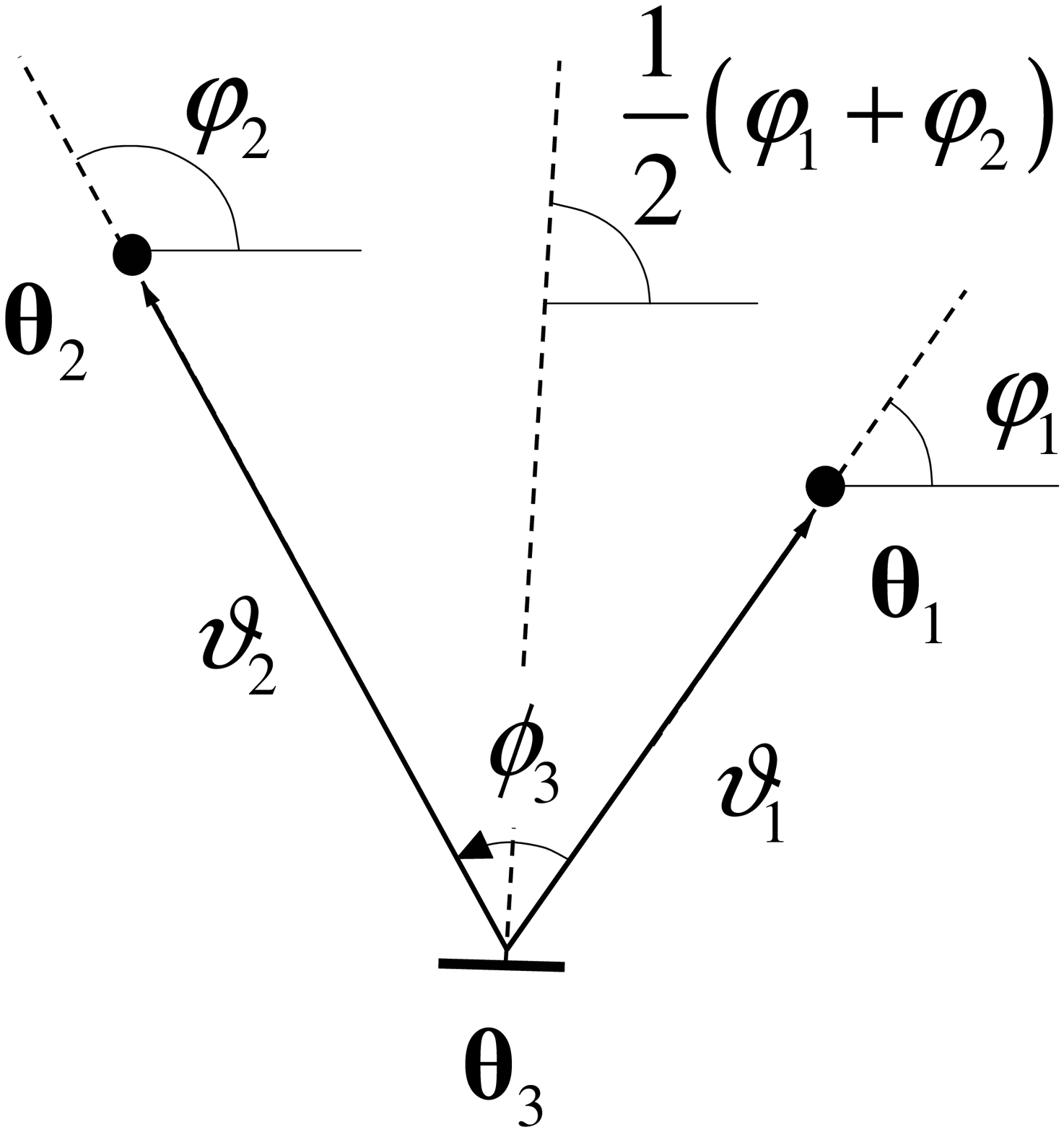,width=6.8cm,clip=}
   \vspace{0.5cm}
   \caption{Geometry of the galaxy-shear-shear correlation, $G_\pm
   \rund{\vt_1,\vt_2,\phi_3}$ (left panel), and the galaxy-galaxy-shear
   correlation, ${\cal G} \rund{\vt_1,\vt_2,\phi_3}$ (right
   panel). Note that the sign of the angle $\phi_3=\vp_2-\vp_1$ is
   important.
              }
         \label{Fig1}
 \end{figure}

\subsection{\llabel{Gddg}Galaxy-shear-shear correlation}
Consider a triplet of points, where two background galaxies are
located at $\vc\theta_1$ and $\vc\theta_2$, and a foreground
galaxy is located at $\vc\theta_3$ (see Fig.\ 1, left). We define
the following galaxy-shear-shear 3PCFs
    \bea
        G_+(\vc\vt_1,\vc\vt_2)&=&G_+(\vt_1,\vt_2,\phi_3)
        =\ave{\gamma(\vc\theta_1;\vp_1)\,\gamma^*(\vc\theta_2;\vp_2)\,\kappa_{\rm
        g}(\vc\theta_3)} \; , \elabel{calGplus} \\
        G_-(\vc\vt_1,\vc\vt_2)&=&G_-(\vt_1,\vt_2,\phi_3)
        =\ave{\gamma(\vc\theta_1;\vp_1)\,\gamma(\vc\theta_2;\vp_2)\,\kappa_{\rm
        g}(\vc\theta_3)} \; , \elabel{calGminus}
    \eea
where the separation vectors between the background galaxies and the
foreground galaxy are $\vc\vt_i=\vc\theta_i-\vc\theta_3$ for
$i=1,2$, and $\vp_i$ is the polar direction of the separation vector
$\vc\vt_i$. These correlation functions depend on the
moduli of the separation vectors $\vc\vt_i$ and the angle $\phi_3$
enclosed by them. In order to find a practical estimator for these
cross-correlations, it is useful to consider the modified correlators
    \be
        \tilde G_+(\vt_1,\vt_2,\phi_3)={1\over \bar N}
        \ave{\gamma(\vc\theta_1;\vp_1)\,\gamma^*(\vc\theta_2;\vp_2)\,N(\vc\theta_3)}
        \; , \;\;
        \tilde G_-(\vt_1,\vt_2,\phi_3)={1\over \bar N}
        \ave{\gamma(\vc\theta_1;\vp_1)\,\gamma(\vc\theta_2;\vp_2)\,N(\vc\theta_3)}
        \;.
    \ee
Using $N/\bar N=1+\kappa_{\rm g}$, these modified correlators become
    \bea
        \tilde G_+(\vt_1,\vt_2,\phi_3)&=&G_+(\vt_1,\vt_2,\phi_3)
        +\ave{\gamma(\vc\theta_1;\vp_1)\,\gamma^*(\vc\theta_2;\vp_2)}
        =G_+(\vt_1,\vt_2,\phi_3)+\xi_+(\Delta\theta) \,{\rm e}^{2{\rm i}\phi_3}\; , \\
        \tilde G_-(\vt_1,\vt_2,\phi_3)&=&G_-(\vt_1,\vt_2,\phi_3)
        +\ave{\gamma(\vc\theta_1;\vp_1)\,\gamma(\vc\theta_2;\vp_2)}
        =G_-(\vt_1,\vt_2,\phi_3)+\xi_-(\Delta\theta) \,
        {\rund{\vt_2{\rm e}^{{\rm i}\phi_3/2}-\vt_1{\rm e}^{-{\rm i}\phi_3/2}}^4
        \over (\Delta\theta)^4} \; ,
        \elabel{GSStilde}
    \eea
where $\Delta\theta=|\vc\vt_2-\vc\vt_1|$ is the separation between
the two background galaxies. The phase factors with which the
shear 2PCFs are multiplied are obtained by rotating the shears
from the direction pointing towards the foreground galaxy to that
of the separation vector of the two background galaxies. If the
polar angle of the latter is denoted by $\vp_\Delta$, then
$\gamma(\vc\theta_i;\vp_i)=\gamma(\vc\theta_i;\vp_\Delta)\, {\rm
e}^{2{\rm i}(\vp_\Delta-\vp_i)}$, from which the expressions for
the phase factors follow. Hence, the modified galaxy-shear-shear
correlators are given by the sum of the (reduced)
galaxy-shear-shear correlators $G_\pm$ (which is proportional to
the cross-bispectrum $b_{\kappa\kappa\rm g}$) plus the 2PCF of
cosmic shear (\ref{eq:XIpm}), modified by a phase factor that
accounts for the fact that the projection directions of the shear
are $\vp_i$, and thus do not correspond to the separation vector between the
background galaxy pair.

We can obtain practical estimators for these cross 3PCFs, as
follows: first, one defines bins in $(\vt_1,\vt_2,\phi_3)$-space.
Then, for each triplet of two background galaxies and one
foreground galaxy whose separation falls inside a given bin, one
adds the product of shears, so that
    \be
        \tilde G_+(\vt_1,\vt_2,\phi_3)={1\over N_{\rm triplet}}\sum_{\rm triplet}
        \gamma(\theta_1;\vp_1)\gamma^*(\theta_2;\vp_2) \; ; \; \; \tilde
        G_-(\vt_1,\vt_2,\phi_3)={1\over N_{\rm triplet}}\sum_{\rm triplet}
        \gamma(\theta_1;\vp_1)\gamma(\theta_2;\vp_2) \; .
    \ee
These estimators can be easily calculated from the catalog of
galaxy positions and ellipticities, e.g., using the method of
Jarvis et al.\ (2004) for finding triplets.  From the estimator
above, the contribution of the shear 2PCF according to
(\ref{eq:GSStilde}) has to be subtracted in order to obtain the
reduced galaxy-shear-shear 3PCF $G_\pm$. Note that observationally
the galaxy-shear-shear correlation is simply the excess of the
shear 2PCF measured around foreground galaxies.

\subsection{\llabel{Gggd}Galaxy-galaxy-shear correlation}
Next we consider the 3PCF between the shear of one background
galaxy and the positions of two foreground galaxies. Let the
former be at position $\vc\theta_3$, the latter at positions
$\vc\theta_1$ and $\vc\theta_2$, then we define the reduced
galaxy-galaxy-shear correlation function as
    \be
        {\cal G}(\vc\vt_1,\vc\vt_2)={\cal G}(\vt_1,\vt_2,\phi_3)
        =\ave{\kappa_{\rm g}(\vc\theta_1)\kappa_{\rm g}(\vc\theta_2)
        \gamma\rund{\vc\theta_3;{\vp_1+\vp_2\over 2}}} \;, \elabel{calG}
    \ee
where, as before, $\vc\theta_i=\vc\theta_3+\vc\vt_i$, and $\vp_i$
is the polar angle of the vector $\vc\vt_i$, for $i=1,2$ (see
Fig.\ 1, right). The shear in (\ref{eq:calG}) is projected along
the line which bisects the angle between the two foreground
galaxies, as measured from the background galaxy. As before, for
obtaining a practical estimator it is useful to first define a
modified 3PCF,
    \be
        \tilde{\cal G}(\vc\vt_1,\vc\vt_2)=\tilde{\cal G}(\vt_1,\vt_2,\phi_3)
        ={1\over \bar N^2}\ave{N(\vc\theta_1)N(\vc\theta_2)
        \gamma\rund{\vc\theta_3;{\vp_1+\vp_2\over 2}}} \;.
        \elabel{calGhat1}
    \ee
Using again $N/\bar N=1+\kappa_{\rm g}$, and applying the
transformation law for rotating shears, this can be written as
    \be
        \tilde{\cal G}(\vc\vt_1,\vc\vt_2)=\tilde{\cal G}(\vt_1,\vt_2,\phi_3)
        ={\cal G}(\vt_1,\vt_2,\phi_3) +\ave{\gamma_{\rm t}}(\vt_1)\,{\rm
        e}^{-{\rm i}\phi_3} +\ave{\gamma_{\rm t}}(\vt_2)\,{\rm e}^{{\rm
        i}\phi_3} \;, \elabel{calGhat2}
    \ee
where the last two terms are the two-point galaxy-shear
correlation functions (\ref{eq:GGL}), obtained from rotating the
shear in the direction of the two foreground galaxies. Hence, the
correlator $\tilde{\cal G}$ is the sum of the reduced
galaxy-galaxy-shear correlation function plus the two GGL
contributions which would be present even for the case of purely
Gaussian density fields. A practical estimator for $\tilde{\cal G}$
is obtained by finding triplets of galaxies that fall in a given
bin, and then summing over the shears of the background galaxies,
    \be
        \tilde{\cal G}={1\over N_{\rm triplet}}\sum_{\rm triplet}
        \gamma\rund{\vc\theta_i;{\vp_1+\vp_2\over 2}} \;.
    \ee
The signature of the galaxy-galaxy-shear correlation is therefore just
the excess shear measured about pairs of foreground
galaxies.

Note that integral measures
of these three-point cross-correlation functions have probably been measured
already. McKay et al.\ (2002) have demonstrated in their measurement of GGL
from the SDSS that the GGL signal is stronger, and extends to much larger
separations, for foreground galaxies that are located in regions of high
galaxy density. This detection provides a correlation between the GGL signal
and the number density of galaxies, and therefore an integral over the 3PCF
${\cal G}$. Furthermore, the galaxy-shear-shear correlation seems to be in the
cosmic shear analysis of the COMBO-17 fields by Brown et al.\ (2003), where
they find a stronger-than-average cosmic shear signal in the A901 field, and a
weaker-than-average cosmic shear signal in the CDFS, which is a field
selected because it is rather poor in brighter galaxies.

The measurement of these galaxy-mass 3PCFs is expected to be
considerably easier than measuring the shear 3PCF itself. Note
that ${\cal G}$ is first order, and $G_\pm$ are second order in
the shear. For the same reason that GGL measurements are easier to
obtain than second-order cosmic shear measurements, the former
being just first order in the shear, one therefore expects that
${\cal G}$ and $G_\pm$ are more straightforward to measure from a
given data set. In particular, every wide-field survey in which
the shear 3PCF can be measured should yield a significant
measurement of these cross-correlation functions.

\section{\llabel{bisp} Relation to the bispectrum}
It is relatively straightforward to write the correlation functions
$G_{\pm}$ and $\cal{G}$ in terms of the (projected) bispectra defined
in Sect.\ts \ref{sc:projected}. The complex shear is related to the
convergence in Fourier space so that $\hat{\gamma_{\rm c}}(\vc \ell) =
\hat{\kappa}(\vc \ell) {\rm e}^{2\rm i \beta}$, where $\beta$ is the
polar angle of the vector $\vc \ell$. Using this, along with the rule
for rotating shears, one can re-write equation (\ref{eq:calGplus})
as
    \be
        G_+(\vc\vt_1,\vc\vt_2) = \int{\d^2\ell_1\over (2\pi)^2}
    \int{\d^2\ell_2\over (2\pi)^2} \;
        {\rm e} ^{-\rm i \rund{\vc\ell_1.\vc\vt_1 + \vc\ell_2.\vc\vt_2}}
    {\rm e} ^{2 \rm i \rund{\beta_1 - \vp_1}}
    {\rm e} ^{2 \rm i \rund{\vp_2 - \beta_2}}
    \, b_{\kappa\kappa{\rm
        g}}(\vc\ell_1,\vc\ell_2;-\vc\ell_1-\vc\ell_2) \; .
    \elabel{FourierInt}
    \ee
Similar relations hold for the correlation functions defined by Eqs.\
(\ref{eq:calGminus}) and (\ref{eq:calG}). Next we split the angles
$\vp_i$ and $\beta_i$ into their mean and their difference, by writing 
$\vp_1 = \zeta - \phi_3/2$,  $\vp_2 = \zeta + \phi_3/2$ and  $\beta_1 =
\eta' - \psi/2$, $\beta_2 = \eta' + \psi/2$, where $\psi$ represents the angle
contained by $\ell_1$ and $\ell_2$, and $\phi_3$ is the same as in
previous sections. 
Then, we write $\vc\ell_1\cdot\vc\vt_1+\vc\ell_2\cdot\vc\vt_2=
A\cos(\eta-\nu)$, where $\eta=\eta'-\zeta$, and
    \be
        A^2 = \ell_1^2 \vt_1^2 + \ell_2^2 \vt_2^2 - 2\ell_1\ell_2 \vt_1
    \vt_2 \cos(\phi_3 - \psi)
    \; , 
\;\;\;
         {\rm e} ^{ 2 \rm i \nu}= \frac{1}{A^2} \left[
     2 \ell_1\ell_2 \vt_1 \vt_2  + \rund{\ell_1\vt_1}^2 {\rm e} ^{\rm i
     (\phi_3-\psi)}
     + \rund{\ell_2\vt_2}^2 {\rm e} ^{-\rm i (\phi_3-\psi)} \right]\;.
    \ee
Splitting up the integrals in (\ref{eq:FourierInt}) into polar 
coordinates in $\ell$-space, the $\eta$-integral can be performed,
yielding 
    \be
        G_+(\vt_1,\vt_2,\phi_3) = \int{\d \ell_1 \, \ell_1\over (2\pi)}
    \int{\d \ell_2 \, \ell_2 \over (2\pi)} \; \int {{\rm d} \psi \over (2\pi)}
    \, \, b_{\kappa\kappa{\rm
        g}}(\ell_1,\ell_2,\psi) \, \,
    {\rm e} ^{- 2 \rm i  \rund{\phi_3 - \psi}} \;  {\rm J}_0(A) \; .
    \elabel{Gplustob}
    \ee
In a similar manner, we find for $G_-$
\bea
        G_-(\vt_1,\vt_2,\phi_3) &=& 
\int{\d^2\ell_1\over(2\pi)^2}
\int{\d^2\ell_2\over(2\pi)^2}\;
{\rm e}^{-{\rm i}(\vc\ell_1\cdot\vc\vt_1+\vc\ell_2\cdot\vc\vt_2)}\,
{\rm e}^{2{\rm i}(\beta_1-\vp_1)}\,
{\rm e}^{2{\rm i}(\beta_2-\vp_2)}\,b_{\kappa\kappa{\rm g}} 
(\vc\ell_1,\vc\ell_2;-\vc\ell_1-\vc\ell_2) \\
&=&
\int{\d \ell_1 \, \ell_1\over (2\pi)}
    \int{\d \ell_2 \, \ell_2 \over (2\pi)} \; 
  \int { {\rm d} \psi\over (2\pi)}
    \, \, b_{\kappa\kappa{\rm
        g}}(\ell_1,\ell_2,\psi) \, \,
    {\rm e} ^{4 \rm i \nu} \; {\rm J}_4(A) \; ,\elabel{Gminustob}
    \eea
and for ${\cal G}$
 \be
        {\cal G}(\vt_1,\vt_2,\phi_3) = \int {\d \ell_1 \, \ell_1\over (2\pi)}
    \int{\d \ell_2 \, \ell_2 \over (2\pi)} \; 
\int { {\rm d} \psi\over (2\pi)}
    \, \, b_{\kappa{\rm g
        g}}(\ell_1,\ell_2,\psi) \, \,
    \frac{\rund{\ell_1{\rm e} ^{-\rm i \psi/2}  + \ell_2 {\rm e} ^{\rm i \psi/2}}^2}{\abs{\ell}^2}\;
    {\rm e} ^{2 \rm i  \nu} \;  {\rm J}_2(A) \; , \elabel{Gtob}
    \ee
with $\abs{\ell}^2 = \ell_1^2 + \ell_2^2 - 2\ell_1\ell_2\cos \psi$.
The above results link directly the cross-correlation functions in
real space to their equivalent projected bispectra. In a similar
fashion, one may also derive a set of inverse relations, i.e
expressing the bispectra in terms of Fourier integrals of the
correlation functions, beginning with the definition of the
bispectrum in Eq.\ (\ref{eq:angbispec}). The resulting
expressions, which are derived in the same way as above, are
\bea
b_{\kappa\kappa{\rm g}}(\vc\ell_1,\vc\ell_2;-\vc\ell_1-\vc\ell_2)
&=&\int\d^2\vt_1 \int\d^2\vt_2 \;
{\rm e}^{{\rm i}(\vc\ell_1\cdot\vc\vt_1+\vc\ell_2\cdot\vc\vt_2)}\,
{\rm e}^{-2{\rm i}(\beta_1-\vp_1)}\,
{\rm e}^{-2{\rm i}(\beta_2-\vp_2)}\,G_-(\vc\vt_1,\vc\vt_2) 
\elabel{btoGminus}\\
&=&\int\d^2\vt_1 \int\d^2\vt_2 \;
{\rm e}^{{\rm i}(\vc\ell_1\cdot\vc\vt_1+\vc\ell_2\cdot\vc\vt_2)}\,
{\rm e}^{-2{\rm i}(\beta_1-\vp_1)}\,
{\rm e}^{2{\rm i}(\beta_2-\vp_2)}\,G_+(\vc\vt_1,\vc\vt_2) \;.
\elabel{btoGplus}\\
\eea
Using the same transformations of the polar angles as before, one more
integration can be carried out, resulting in
\bea
b_{\kappa\kappa{\rm g}}(\ell_1,\ell_2,\psi)
&=&2\pi\int\d\vt_1\;\vt_1\int\d\vt_2\;\vt_2\int\d\phi_3\;
{\rm e}^{4{\rm i}\nu}\,{\rm J}_4(A)\;G_-(\vt_1,\vt_2,\phi_3)
\\
&=&2\pi\int\d\vt_1\;\vt_1\int\d\vt_2\;\vt_2\int\d\phi_3\;
{\rm e}^{2{\rm i}(\psi-\phi_3)}\,{\rm J}_0(A)\;
G_+(\vt_1,\vt_2,\phi_3) \;.
\eea
In a similar way, we can express the cross-bispectrum $b_{{\rm
gg}\kappa}$ in terms of ${\cal G}$, yielding
\be
b_{{\rm gg}\kappa}(\ell_1,\ell_2,\psi)
=2\pi\int\d\vt_1\;\vt_1\int\d\vt_2\;\vt_2\int\d\phi_3\;
\frac{\rund{\ell_1{\rm e} ^{\rm i \psi/2}  + \ell_2 {\rm e} ^{-\rm i \psi/2}}^2}{\abs{\ell}^2}\;
    {\rm e} ^{-2 \rm i  \nu} \;  {\rm J}_2(A) \;{\cal G}(\vt_1,\vt_2,\phi_3)
\; .
    \ee
Using (\ref{eq:FourierInt}) and replacing the cross-bispectrum in
favour of $G_-$, using (\ref{eq:btoGminus}), one can express $G_+$ in
terms of $G_-$. Similarly, using (\ref{eq:Gminustob}) and
(\ref{eq:btoGplus}), we obtain the inverse relation; these read:
\bea
G_+(\vt_1,\vt_2,\phi_3)&=&{2\over \pi}\int\d\vt\;\vt\int\d\phi\;
{ \eck{\vt\,{\rm e}^{{\rm i}(\phi_3-\phi)/2} -
\vt_2\,{\rm e}^{-{\rm i}(\phi_3-\phi)/2} }^4\over
\eck{\vt^2+\vt_2^2-2\vt\vt_2\cos(\phi_3-\phi)}^3 }\;
G_-(\vt_1,\vt,\phi) \;;
\\
G_-(\vt_1,\vt_2,\phi_3)&=&{2\over \pi}\int\d\vt\;\vt\int\d\phi\;
{ \eck{\vt\,{\rm e}^{-{\rm i}(\phi_3-\phi)/2} -
\vt_2\,{\rm e}^{{\rm i}(\phi_3-\phi)/2} }^4\over
\eck{\vt^2+\vt_2^2-2\vt\vt_2\cos(\phi_3-\phi)}^3 }\;
G_+(\vt_1,\vt,\phi) \;.
\eea
Hence, if the shear field is a pure E-mode field, these interrelations
will be satisfied; note that similar equations are valid for the shear
2PCF (Crittenden et al.\ 2002; Schneider et al.\ 2002), relating
$\xi_+$ to $\xi_-$ and vice versa.

\def\com#1{{\breve#1}}

\section{\llabel{Aperture}Aperture statistics}
In Sect.\ts 5 we defined 3PCFs for the galaxy-mass
correlations, and gave practical estimators for measuring them.
These 3PCFs were then related to the corresponding bispectra through
Fourier transform relations in Sect.\ts 6. As was demonstrated in
Schneider et al.\ 
(2004), the relation between the shear 3PCF and the underlying
projected mass bispectrum is rather complicated and not easy to
evaluate. From Eqs.\ (\ref{eq:Gplustob}) through (\ref{eq:Gtob}) we
observe that the same is true for the relation between the
correlation functions considered here and the corresponding
bispectra.

However, we can avoid the use of numerically complicated
transformations between the 3PCF and the bispectra by considering
aperture statistics (Schneider 1996, 1998; van Waerbeke 1998;
Crittenden et al.\ 2002), in the same way as has been done for the
shear 3PCF (Jarvis et al.\ 2004; Schneider et al.\ 2004) and for
the measurement of the second-order bias factor and galaxy-mass
correlation coefficient (Hoekstra et al.\ 2001, 2002). The main
reason for this is that the third-order aperture statistics on the one hand
provide a very localized measurement of the
corresponding bispectra, and on the other hand can be readily
evaluated from the corresponding 3PCFs. In this section, we apply
the aperture statistics to the galaxy-mass 3PCFs. The aperture
mass is defined as
    \be
        M_{\rm ap}(\theta)=\int
        \d^2\vt\;U_\theta(|\vc\vt|)\,\kappa(\vc\vt) =\int
        \d^2\vt\;Q_\theta(|\vc\vt|)\,\gamma_{\rm t}(\vc\vt)\;, {\rm with}
        \;\;
        Q_\theta(\vt)={2\over\vt^2}\int_0^\vt\d\vt'\,\vt'\,U_\theta(\vt')
        -U_\theta(\vt) \;, \elabel{Mapdef}
    \ee
where $U_\theta$ is a weight function of zero total weight, i.e.,
$\int\d\vt\,\vt\,U_\theta(\vt)=0$. The scale of the weight
function is described by $\theta$, and $\gamma_{\rm t}$ is the
tangential shear component as measured with respect to the
direction towards the center of the aperture. Furthermore, we
define the aperture counts
    \be
        {\cal N}(\theta)=\int \d^2
        \vt\;U_\theta(|\vc\vt|)\,\kappa_{\rm g}(\vc\vt) ={1\over \bar
        N}\int \d^2 \vt\;U_\theta(|\vc\vt|)\,N(\vc\vt) \;,
    \ee
where the final equality follows because $U$ is a compensated
filter function. We consider here the filter function introduced
by Crittenden et al.\ (2002), which has also been used by Jarvis
et al.\ (2004) and Schneider et al.\ (2004). Writing
$U_\theta(\vt)=\theta^{-2}\,u(\vt/\theta)$, the filter
function reads
    \be
        u(x)={1\over 2\pi}\rund{1-{x^2\over 2}}\,{\rm
        e}^{-x^2/2}\; ;\;\; \hat u(\eta)=\int \d^2 x\;u(|\vc x|)\,{\rm
        e}^{{\rm i}\vc\eta\cdot\vc x} ={\eta^2\over 2}\,{\rm
        e}^{-\eta^2/2}\; ; \;\; Q_\theta(\vt)={\vt^2\over
        4\pi\theta^4}\exp\rund{-{\vt^2\over 2\theta^2}}\;.
        \elabel{u+uFour}
    \ee
This choice for $U_\theta$ has the disadvantage that the support
of the filter is formally infinite; however, the Gaussian factor
renders the {\em effective} range of support finite, since the
filter function becomes extremely small for $\vt\gtrsim 3\theta$.
This disadvantage over other filter functions that have been
employed in cosmic shear studies (Schneider et al.\ 1998) is more
than compensated by the convenient mathematical properties, of
which we will make extensive use below.

We can now relate the third-order correlations between the
aperture measures to the corresponding bispectra. Consider first
    \[
        \ave{M_{\rm ap}(\theta_1)M_{\rm ap}(\theta_2){\cal N}(\theta_3)}
        \!\equiv \! \ave{M_{\rm ap}M_{\rm ap}{\cal
        N}}(\theta_1,\theta_2;\theta_3)\!\!=\!\!
        \int\!\!\d^2\vt_1\;U_{\theta_1}(\vc\vt_1)
        \!\!\int\!\!\d^2\vt_2\;U_{\theta_2}(\vc\vt_2)
        \!\!\int\!\!\d^2\vt_3\;U_{\theta_3}(\vc\vt_3)
        \ave{\kappa(\vc\vt_1)\kappa(\vc\vt_2)\kappa_{\rm g}(\vc\vt_3)} \;.
    \]
Replacing the $\kappa_i$ by their Fourier transforms, carrying out
the $\vt_i$-integrations, making use of (\ref{eq:u+uFour}) and the
definition of the bispectra, yields
    \be
        \ave{M_{\rm ap}M_{\rm ap}{\cal N}}(\theta_1,\theta_2;\theta_3)=
        \int{\d^2\ell_1\over (2\pi)^2}\int{\d^2\ell_2\over (2\pi)^2} \;
        \hat u(\ell_1\theta_1)\,\hat u(\ell_2\theta_2)\,
        \hat u(|\vc\ell_1+\vc\ell_2|\theta_3)\, b_{\kappa\kappa{\rm
        g}}(\vc\ell_1,\vc\ell_2;-\vc\ell_1-\vc\ell_2) \;.
    \ee
Similarly, we obtain for the galaxy-galaxy-mass aperture correlator
    \be
        \ave{{\cal N}{\cal N}M_{\rm ap}}(\theta_1,\theta_2;\theta_3)=
        \int{\d^2\ell_1\over (2\pi)^2}\int{\d^2\ell_2\over (2\pi)^2} \;
        \hat u(\ell_1\theta_1)\,\hat u(\ell_2\theta_2)\,
        \hat u(|\vc\ell_1+\vc\ell_2|\theta_3)\, b_{{\rm
        gg}\kappa}(\vc\ell_1,\vc\ell_2;-\vc\ell_1-\vc\ell_2) \;.
    \ee
Since $\hat u$ is a function that has a very narrow peak, the
third-order aperture measures provide very localized information
on the respective bispectra and are thus ideal for probing the
latter. Unless the bispectra have very sharp features, the
third-order aperture measures contain essentially all information
about the bispectra -- cf. the corresponding discussion for the
shear 3PCF and their aperture measures in Schneider et al.\
(2004). We next show how the third-order aperture measures can be
calculated directly in terms of the respective correlation
functions. For that, it is convenient to introduce a complex
aperture shear measure, defined as
    \be
        M(\theta):=M_{\rm ap}(\theta)+{\rm i}M_\perp(\theta)
        =\int \d^2\vt\;Q_\theta(|\vc\vt|)\eck{\gamma_{\rm t}(\vc\vt)+{\rm
        i}\gamma_\times(\vc\vt)}=\int \d^2\vt\;Q_\theta(|\vc\vt|)
        \,\gamma(\vc\vt;\vp)
        \elabel{Mdef}
    \ee
where $\vp$ is the polar angle of $\vc\vt$. If the shear field is entirely due
to the lensing mass distribution, $M_\perp$ vanishes identically (Crittenden
et al.\ 2002). A non-zero value for $M_\perp$ would indicate the presence of
B-mode shear. In fact
$M_{\rm ap}(\theta)$ vanishes identically for B-modes, whereas
$M_\perp(\theta)$ yields zero for a pure E-mode field. Thus, the
aperture measures are ideally suited to separating E- and B-modes of
the shear.

\subsection{Galaxy-mass-mass aperture statistics}
We will now express the third-order aperture measures to the 3PCFs
considered in the previous section. First we find from the
definition (\ref{eq:Mdef})
    \be
        \ave{MM{\cal N}}(\theta_1,\theta_2;\theta_3)=
        \int\d^2 X_1 \int\d^2 X_2 \int\d^2 Y\;
        Q_{\theta_1}(|\vc X_1|) Q_{\theta_2}(|\vc X_2|) U_{\theta_3}(|\vc Y|)
        \;
        \ave{\gamma(\vc X_1;\psi_1)\gamma(\vc X_2;\psi_2)\kappa_{\rm g}(\vc Y)} \;,
        \elabel{MMN1}
    \ee
where $\psi_i$ is the polar angle of the vector $\vc X_i$, $i=1,2$. Next we
introduce the separation vectors $\vc\vt_i=\vc X_i- \vc Y$, and let $\vp_i$
denote their polar angle. Then we apply the transformation
   \be
        \gamma(\vc X_i;\psi_i)=\gamma(\vc Y+\vc\vt_i;\vp_i)\,{\rm e}^{2{\rm
        i}(\vp_i-\psi_i)}
        =\gamma(\vc Y+\vc\vt_i;\vp_i)\,{\rm e}^{2{\rm i}\vp_i}\,
        {\rund{\com Y^*+\com\vt_i^*}^2\over \abs{\vc Y+\vc \vt_i}^2} \;,
    \ee
where we have introduced the notation that a two-dimensional vector $\vc X$
can be written as a complex number $\com X:=X_1 +{\rm i} X_2$, which is
convenient for expressing phases: If $\vp$ is the polar angle of $\vc X$, then
${\rm e}^{{\rm i}\vp} = \com X/|\vc X|$.  Inserting the foregoing expression
into (\ref{eq:MMN1}) and using the definition (\ref{eq:u+uFour}) of the
filter functions, one obtains
    \bea
        \ave{MM{\cal N}}(\theta_1,\theta_2;\theta_3)&=&
        {1\over (4\pi)^3\,\theta_1^4\theta_2^4\theta_3^4}
        \int \d^2\vt_1\;{\rm e}^{2{\rm i}\vp_1}
        \int \d^2\vt_2\;{\rm e}^{2{\rm i}\vp_2}
        \;G_-(\vc\vt_1,\vc\vt_2) \nonumber \\
        \times
        \int\d^2 Y \!\!\!\!&&\!\!\!\!
        \rund{\com Y^*+\com\vt_1^*}^2\rund{\com Y^*+\com\vt_2^*}^2
        \rund{2\theta_3^2-\abs{\vc Y}^2}\,
        \exp\eck{-\rund{ {\abs{\vc Y + \vc\vt_1}^2\over 2\theta_1^2}
        +{\abs{\vc Y + \vc\vt_2}^2\over 2\theta_2^2}
        +{\abs{\vc Y}^2\over 2\theta_3^2} } } \;.
        \elabel{MMN2}
    \eea
The $\vc Y$-integration can be performed, and the result, when
multiplied with the phase factors, only depends on the modulus of
the $\vc\vt_i$ and the angle $\phi_3$ they enclose. Hence,
    \be
        \ave{MM{\cal N}}(\theta_1,\theta_2;\theta_3)= \int_0^\infty
        \d\vt_1\;\vt_1\int_0^\infty\d\vt_2\;\vt_2\int_0^{2\pi}
        \d\phi_3\;G_-(\vt_1,\vt_2,\phi_3)\, A_{MM{\cal
        N}}(\vt_1,\vt_2,\phi_3|\theta_1,\theta_2;\theta_3) \;,
        \elabel{MMN3}
    \ee
where the convolution function $A_{MM{\cal N}}$ is given in the
appendix. In the same way, we can evaluate
    \bea
        \ave{MM^*{\cal
        N}}(\theta_1,\theta_2;\theta_3)&=& {1\over
        (4\pi)^3\,\theta_1^4\theta_2^4\theta_3^4} \int \d^2\vt_1\;
        {\rm e}^{2{\rm i}\vp_1}
        \int \d^2\vt_2\;
        {\rm e}^{-2{\rm i}\vp_2}
        \;G_+(\vc\vt_1,\vc\vt_2) \nonumber \\
        &\times&
        \int\d^2 Y \rund{\com Y^*+\com\vt_1^*}^2\rund{\com Y+\com\vt_2}^2
        \rund{2\theta_3^2-\abs{\vc Y}^2}\,
        \exp\eck{-\rund{ {\abs{\vc Y + \vc\vt_1}^2\over 2\theta_1^2}
        +{\abs{\vc Y + \vc\vt_2}^2\over 2\theta_2^2}
        +{\abs{\vc Y}^2\over 2\theta_3^2} } }
        \elabel{MMsN}
        \\
        &=&
        \int_0^\infty \d\vt_1\;\vt_1\int_0^\infty\d\vt_2\;\vt_2\int_0^{2\pi}
        \d\phi_3\;G_+(\vt_1,\vt_2,\phi_3)\,
        A_{MM^*{\cal N}}(\vt_1,\vt_2,\phi_3|\theta_1,\theta_2;\theta_3) \;,
    \elabel{MMsN1}
    \eea
where again the convolution function is given in the appendix. By combining
(\ref{eq:MMN3}) and (\ref{eq:MMsN}), one finds
    \bea
        \ave{M_{\rm ap}M_{\rm ap}{\cal
        N}}(\theta_1,\theta_2;\theta_3)
        &=& \Re\left[\ave{MM{\cal N}}(\theta_1,\theta_2;\theta_3) + \ave{MM^*{\cal
        N}}(\theta_1,\theta_2;\theta_3)\right]/2 \nonumber \\
        \ave{M_\perp M_\perp{\cal
        N}}(\theta_1,\theta_2;\theta_3)
        &=& \Re\left[\ave{MM^*{\cal N}}(\theta_1,\theta_2;\theta_3) - \ave{MM{\cal
        N}}(\theta_1,\theta_2;\theta_3)\right]/2 \nonumber \\
        \ave{M_\perp M_{\rm ap}{\cal
        N}}(\theta_1,\theta_2,\theta_3)
        &=& \Im\left[\ave{MM{\cal N}}(\theta_1,\theta_2;\theta_3) + \ave{MM^*{\cal
        N}}(\theta_1,\theta_2;\theta_3)\right]/2 \;.\nonumber
    \eea
The first of these is the expression that is directly related to
the bispectrum $b_{\kappa\kappa{\rm g}}$. The second expression is
expected to vanish if the shear is caused solely by gravitational
lensing. A significant non-zero value of this correlator would
indicate the presence of B-mode shear which is correlated with the
foreground galaxy distribution. The final expression is expected
to be zero because it is not parity-invariant, as all odd-order
statistics in B-mode shear (Schneider 2003); hence, a measured
non-zero value would indicate that the data violate parity
invariance.

One may consider how the results defined above would change if the modified
correlators $\tilde{G}_{\pm}$ were used in the definition of $\ave{MM^*{\cal
N}}$ and $\ave{MM{\cal N}}$. Using the fact that $N/\bar{N} = \kappa_{\rm g} + 1$ we
find, using Eq.\ (\ref{eq:MMN1}), that since $U$ is a compensated filter,
the additional terms arising from two point shear-shear correlations
vanish. Thus we make the following useful assertion that when working with
the aperture mass statistics, one need not subtract the contribution from the
shear 2PCF to compute the reduced three point
statistics.

\subsection{Galaxy-galaxy-mass aperture statistics}
Finally, we consider the galaxy-galaxy-mass third-order aperture
statistics
    \be
        \ave{{\cal N}{\cal N}M}(\theta_1,\theta_2;\theta_3)=
        \int\d^2 X_1 \int\d^2 X_2 \int\d^2 Y\;
        U_{\theta_1}(|\vc X_1|) U_{\theta_2}(|\vc X_2|) Q_{\theta_3}(|\vc Y|)
        \;
        \ave{\kappa_{\rm g}(\vc X_1) \kappa_{\rm g}(\vc X_2) \gamma(\vc Y;\psi)}\;,
\elabel{NNM1}
\ee
where $\psi$ is the polar angle of $\vc Y$. Again introducing the separation
vectors $\vc\vt_i=\vc X_i -\vc Y$, which have polar angles $\vp_i$, $i=1,2$,
using the transformation
    \be
        \gamma(\vc Y;\psi)=\gamma\rund{\vc Y;{\vp_1+\vp_2\over 2}}\,
        {\rm e}^{{\rm i}(\vp_1+\vp_2-2\psi)}
        =\gamma\rund{\vc Y;{\vp_1+\vp_2\over 2}}\,{\rm e}^{{\rm i}\vp_1}\,{\rm
        e}^{{\rm i}\vp_2}\,{(\com Y^*)^2\over |\vc Y|^2}\;,
    \ee
and inserting the definitions (\ref{eq:u+uFour}) into
(\ref{eq:NNM1}) yields
    \bea \ave{{\cal N}{\cal
        N}M}(\theta_1,\theta_2;\theta_3)&=& {1\over
        (4\pi)^3\,\theta_1^4\theta_2^4\theta_3^4} \int \d^2\vt_1\; {\rm
        e}^{{\rm i}\vp_1} \int \d^2\vt_2\; {\rm e}^{{\rm i}\vp_2}
        \;{\cal G}(\vc\vt_1,\vc\vt_2) \nonumber \\
        \times
        \int\d^2 Y \!\!\!\!  &&  \!\!\!\!
        \rund{2\theta_1^2-\abs{\vc Y+\vc\vt_1}^2}
        \rund{2\theta_2^2-\abs{\vc Y+\vc\vt_2}^2}\,\rund{\com Y^*}^2
        \exp\eck{-\rund{ {\abs{\vc Y + \vc\vt_1}^2\over 2\theta_1^2}
        +{\abs{\vc Y + \vc\vt_2}^2\over 2\theta_2^2}
        +{\abs{\vc Y}^2\over 2\theta_3^2} } }
        \elabel{NNM2}    \\
        &=&
        \int_0^\infty \d\vt_1\;\vt_1\int_0^\infty\d\vt_2\;\vt_2\int_0^{2\pi}
        \d\phi_3\;{\cal G}(\vt_1,\vt_2,\phi_3)\,
        A_{{\cal NN}M}(\vt_1,\vt_2,\phi_3|\theta_1,\theta_2;\theta_3) \;,
    \elabel{NNM3}
    \eea
where the convolution function is also derived in the appendix. The real part
of $\ave{{\cal N}{\cal N}M}$ corresponds to the aperture 3PCF that is directly
related to the bispectrum $b_{{\rm gg}\kappa}$, whereas the imaginary part
vanishes due to parity invariance. Finally in this section we note that, as
with the galaxy-mass-mass statistics, the quantity $\ave{{\cal N}{\cal N}M}$
does not change when the definition for
$\tilde{\cal{G}}$ is used in place of $\cal{G}$ in Eq.\
(\ref{eq:NNM1}). Second-order galaxy-galaxy lensing effects are
therefore not important for 
the three-point aperture mass statistics.

\section{Discussion}
In this paper, we have defined third-order galaxy-mass correlation
functions and the corresponding third-order bias factor and
related them to quantities which can be measured directly through
galaxy-galaxy-galaxy lensing. Whereas the underlying physical
quantities are the bispectra of the mass and galaxy distribution,
as well as the cross-bispectra, the observables are the shear as
measured from the ellipticity of background galaxies in relation
to the position of foreground galaxies. We have argued that these
third-order correlations have probably been measured already, and
are almost certainly measurable from data sets existing now; the
ongoing and planned wide-field surveys will obtain precision
measurements of these third-order statistics. The basic
observables are the 3PCFs, as their measurement is not affected by
holes and gaps in the data set which are unavoidable due to the
selection of the survey geometry and masking of bright stars and
galaxies, as well as CCD defects. The 3PCFs are related to the
corresponding bispectra through a very broad and oscillating
filter function (see Schneider et al.\ 2004 for the case of the
shear 3PCF) and thus do not directly provide information about the
shape of the bispectra. Therefore, we have considered the aperture
statistics, which on the one hand can be calculated readily from
the measured correlation functions, and on the other hand are
related to the bispectra through a narrow filter function, hence
providing localized information on the latter. In fact, unless the
bispectra vary strongly as a function of their angular wave
vectors $\vc\ell_i$, the third-order aperture measures are
expected to contain essentially all information about the
third-order galaxy-mass correlations. Furthermore, the aperture
measures allow one to easily identify the presence of B-modes in
the shear field, as well as unphysical parity-violating
contributions, which can only be due to the observing and data
reduction steps, unless we drop the assumption that the mass
distribution in our Universe is parity invariant.

The convolution functions relating the 3PCFs to the third-order
aperture measures have in principle an infinite support; in practice,
however, the exponential factor implies that the 3PCFs have to be
measured only up to a few times the corresponding filter scales
$\theta_i$. Therefore, these convolution functions have
essentially finite support. 

Nevertheless, in order to calculate the
third-order aperture measures, the corresponding correlation functions
need to be measured to $\sim 6\theta_{\rm max}$, where $\theta_{\rm
max}$ is the largest of the three aperture radii. The largest scale on
which the correlation functions can be measured is determined by the
geometry of the survey. For a large survey with contiguous area, this
scale is limited by the smallness of the signal when entering the
linear regime of structure evolution. In this case, the aperture
measures contain all the information contained in the measured
correlation functions. If the survey consists of independent patches
each of size $\psi$, then the correlation functions can be measured
for angular scales up to $\sim\psi$, meaning that the aperture measures
are limited to $\theta_{\rm max}\lesssim \psi/6$. In this case, the
measured correlation functions contain more information than the
aperture measures. In a future work, we plan to study these aspects;
note, however, that estimates of the `information content' are
tedious, since they involve the covariance of third-order statistical
measures, and thus depend, in general, on six variables and have to be
estimated from sixth-order statistics.

The correlation function ${\cal G}$ measures the same cross-bispectrum as the
third-order statistics considered by M\'enard et al.\ (2003). In that paper, they
considered the correlation between pairs of foreground galaxies and background
QSOs. Due to the magnification of distant QSOs by the matter in which the
foreground galaxies are embedded, and the steep slope of bright QSO source
counts, such a correlation should be measurable with the large QSO sample
obtained by the Sloan Digital Sky Survey. Instead of magnification, we employ
the shear around galaxy pairs. Which of the two methods is better able to
measure the galaxy-mass bispectrum depends on the available observational
data. For deep wide-field images, the shear method investigated here will be
more efficient, given the sparseness of bright QSOs on the sky.

As is true for the second-order galaxy-mass correlations, as
measured in GGL, the physical interpretation of the third-order
correlations is not straightforward, but needs to be done in the
frame of a model. On very large scales, we might expect that $b_3$
essentially becomes a constant, but that will be difficult to
measure, as on these large scales, the density field is expected to
quickly approach a Gaussian, and thus the bispectrum should be very
small.

A useful analytic description of clustering statistics on all cosmological
scales is given by the halo model (see Cooray \& Sheth \ 2002 for a review).
Within this framework it will be possible to relate the higher-order
cross-correlation functions to the quantities that specify the Halo Occupation
Distribution.  As pointed out by Berlind \& Weinberg (2002) the HOD
essentially contains {\em all} of the information about the statistics of
galaxy clustering that theories of galaxy formation are able to provide; it is
a complete description of the bias. Future empirical determinations of the HOD
using observations of galaxy and mass clustering will enable powerful
constraints to be placed on the development of models for galaxy formation. In
a forthcoming paper we shall explore these concepts in detail, building the
physical interpretation of our third-order galaxy-mass correlations using the
halo model and the HOD.

\begin{acknowledgement}
We thank Martin Kilbinger for useful comments on the manuscript.
This work was supported by the German Ministry for
Science and Education (BMBF) through the DLR under the project 50 OR
0106 and by the Deutsche
Forschungsgemeinschaft under the project SCHN 342/3--1.

\end{acknowledgement}

\appendix

\section{The convolution functions}
In this appendix, we derive the convolution functions which relate the
third-order aperture measures to the corresponding 3PCFs. We start by rewriting
the exponent in (\ref{eq:MMN2}), (\ref{eq:MMsN}) and (\ref{eq:NNM2}) by
making a translation $\vc Y=\vc y-\vc c$, where $\vc c$ is chosen so
as to remove
linear terms in $\vc y$,
    \be
        {\abs{\vc Y + \vc\vt_1}^2\over 2\theta_1^2}
        +{\abs{\vc Y + \vc\vt_2}^2\over 2\theta_2^2}
        +{\abs{\vc Y}^2\over 2\theta_3^2}={\abs{\vc y}^2\over a_2}+b_0 \;; \;\;
        a_2={2\theta_1^2\theta_2^2\theta_3^2 \over
        \theta_1^2\theta_2^2+\theta_1^2\theta_3^2+\theta_2^2\theta_3^2} \; ; \;\;
        \vc c={a_2\over 2}\rund{ {\vc\vt_1\over \theta_1^2}+{\vc\vt_2\over
        \theta_2^2}} \; ;
    \ee
    \be
        b_0={\abs{\vc\vt_1}^2\over 2\theta_1^2}+{\abs{\vc\vt_2}^2\over 2\theta_2^2}
        -{\abs{\vc c}^2\over a_2}=
        {\vt_1^2\over 2\theta_1^2}+{\vt_2^2\over 2\theta_2^2}
        -{a_2\over 4}\rund{ {\vt_1^2\over \theta_1^4} +
        {2\vt_1\vt_2\cos\phi_3\over\theta_1^2\theta_2^2}
        +{\vt_2^2\over \theta_2^4}} \;.
    \ee
Next, we consider the prefactor of the exponential in
(\ref{eq:MMN2}), and change the integration variable from $\vc Y$
to $\vc y$. For this, we first define
    \be
        \rund{\com Y^* + \com\vt_i^* } {\rm e}^{{\rm i}\vp_i} =\com y^* {\rm e}^{{\rm
        i}\vp_i} +\vt_i - \com c^* {\rm e}^{{\rm i}\vp_i} \equiv \com y^*
        {\rm e}^{{\rm i}\vp_i} +\com g_i \;, \;\;
        \vt_i\equiv\abs{\vc\vt_i}\;, \ee for $i=1,2$, or more explicitly,
        \be \com g_1=\vt_1-{a_2\over 2}\rund{{\vt_1\over \theta_1^2}
        +{\vt_2\,{\rm e}^{-{\rm i}\phi_3} \over \theta_2^2}} \; ; \;\;
        \com g_2=\vt_2-{a_2\over 2}\rund{{\vt_2\over \theta_2^2}
        +{\vt_1\,{\rm e}^{{\rm i}\phi_3} \over \theta_1^2}} \; .
    \ee
The prefactor of the exponential, when multiplied with the phase
factors, then becomes
    \bea
        F'_{MM{\cal N}} &=& \rund{\com y^* {\rm
        e}^{{\rm i}\vp_1} +\com g_1}^2 \rund{\com y^* {\rm e}^{{\rm
        i}\vp_2} +\com g_2}^2 \rund{2\theta_3^2-\abs{\vc y}^2-\abs{\vc
        c}^2+\com c \com y^*+\com c^* \com y}
        \nonumber \\
        &=&\com g_1^2\com g_2^2 \rund{2\theta_3^2-\abs{\vc y}^2-\abs{\vc c}^2}
        +2\abs{\vc y}^2\com g_1\com g_2 \eck{\com g_2\rund{\vt_1-\com g_1}
        +\com g_1\rund{\vt_2-\com g_2}} + C \; = F_{MM{\cal N}} + C,
    \eea
where $C$ denotes additional terms which, however, become zero
when the $\vc y$-integration is carried out, since the exponent
depends only on $\abs{\vc y}^2$. The $\vc y$-integration can now
be carried out,
    \[
        \int\d^2 y \, F'_{MM{\cal N}}\, {\rm e}^{-\abs{\vc y}^2/a_2}
        = 2\pi\int_0^\infty \!\!\d y\,y\,F_{MM{\cal N}}\,{\rm e}^{-\abs{\vc y}^2/a_2}
        =\pi a_2^2 \com g_1 \com g_2
        \eck{\rund{ {\theta_3^2\over\theta_1^2} +{\theta_3^2\over\theta_2^2}
        -{\abs{\vc c}^2\over a_2}}\com g_1 \com g_2
        +2\rund{\com g_2 \vt_1 + \com g_1 \vt_2-2\com g_1\com g_2}}\; .
    \]
The result depends only on $\vt_1$, $\vt_2$ and the angle
$\phi_3$ between $\vc\vt_1$ and $\vc\vt_2$, so that an additional
angular integration can be carried out in Eq.\ (\ref{eq:MMN2}). We
thus arrive at (\ref{eq:MMN3}), with
    \be
        A_{MM{\cal N}}= {\com g_1 \com g_2\,{\rm e}^{-b_0} \over 72\pi\Theta^8}
        \;\eck{\rund{ {\theta_3^2\over\theta_1^2} +{\theta_3^2\over\theta_2^2}
        -{\abs{\vc c}^2\over a_2}}\com g_1 \com g_2
        +2\rund{\com g_2\vt_1 + \com g_1\vt_2-2\com g_1\com g_2}} \; ;
        \;\;
        \Theta^4={\theta_1^2\theta_2^2+\theta_1^2\theta_3^2+\theta_2^2\theta_3^2 \over
        3} \;.
    \ee

Similar expansions of the prefactors of the exponentials in
Eqs.\ (\ref{eq:MMsN}) and (\ref{eq:NNM2}) can also be
determined using the same definitions as above. Performing the
relevant integrals, we find that the convolution factors for
equations (\ref{eq:MMsN1}) and (\ref{eq:NNM3}) are then given by
    \bea
        A_{MM^*{\cal N}} & = & {{\rm e}^{-b_0} \over 72\pi\Theta^8}
    \Big[2\rund{\vt_1 \com g^*_2 + \vt_2 \com g_1 -2 \com g_1 \com g^*_2}
    \rund{\com g_1\com g^*_2 + 2a_2 {\rm e}^{- {\rm i} \phi_3}}
     + \,  2 a_2 \rund{2\theta_3^2 -\abs{\vc c}^2 - 3a_2} {\rm e}^{-2 {\rm i} \phi_3} \nonumber \\
     & & \hspace{5cm}
     + \, 4 \com g_1 \com g^*_2 \rund{2\theta_3^2 -\abs{\vc c}^2 - 2a_2} {\rm e}^{{-\rm i} \phi_3}
     + \, \frac{\rund{\com g_1 \com g^*_2}^2}{a_2}\rund{2\theta_3^2 -\abs{\vc c}^2 - a_2}\Big]
    \eea
and
    \bea
        A_{{\cal N}{\cal N}M}  =  {{\rm e}^{-b_0} \over 72\pi\Theta^8}
  \!\!\!&&\!\!\!
	\Big\{ \rund{\com g_1 -\vt_1 } \rund{\com g_2 - \vt_2}\Big[\, 
	\frac{1}{a_2}F_1 F_2 - \rund{F_1 + F_2} + 2a_2 + \com g_1^* \com g_2
        {\rm e}^{- {\rm i} \phi_3} +  \com g_1\com g^*_2 {\rm e}^{ {\rm i} \phi_3} 
	 \Big] \nonumber \\
    & & \vspace{0.5cm} - \left[ \rund{\com g_2 -
        \vt_2} + \rund{\com g_1 - \vt_1} {\rm e}^{{\rm i} \phi_3}\right]
	\left[\com g_1\rund{F_2-2a_2} + \com g_2 \rund{F_1-2a_2} {\rm e}^{- {\rm i} \phi_3}\right]
	 +  2 \com g_1 \com g_2 a_2  \Big\} \, 
    \eea
where $F_i = 2\theta_i^2 - \abs{\com g_i}^2$.


\begin{thebibliography}{}

\bibitem[2000]{BRE00}
  Bacon, D.J., Refregier, A.R. \& Ellis, R.S. 2000, MNRAS, 318, 625
\bibitem[2001]{BS01}
  Bartelmann M., Schneider P., 2001, Phys. Rep., 340, 291 (BS01)
\bibitem[2002]{BerlindWeinberg}
Berlind A. \& Weinberg, D.\ 2002, ApJ 575, 587
\bibitem[2002]{BMvW02}
  Bernardeau, F., Mellier, Y. \& van Waerbeke, L.\ 2002, A\&A 389, L28
\bibitem[1997]{BvWM97}
  Bernardeau, F., van Waerbeke, L. \& Mellier, Y.\ 1997, A\&A, 322, 1
\bibitem[1991]{BSBV}
  Blandford, R.D., Saust, A.B., Brainerd, T.G. \& Villumsen,
  J.V. 1991, MNRAS, 251, 600
\bibitem[1996]{BBS}
  Brainerd, T.G., Blandford, R.D., \& Smail, I.\ 1996, ApJ, 466, 623
\bibitem[2003]{Brown03}
  Brown, M.L., Taylor, A.N., Bacon, D.J.\ et al.\ 2003, MNRAS 341, 100
\bibitem[2002]{Cooraysheth}
Cooray, A. \& Sheth, R.\ 2002, PhR 372, 1
\bibitem[2002]{critt02}
  Crittenden, R.G., Natarajan, P., Pen, U.-L. \& Theuns, T. 2002,
  ApJ, 568, 20
\bibitem[2001]{crometz}
  Croft, R.A.C. \& Metzler, C.A.\ 2001, ApJ, 545, 561
\bibitem[1996]{DellTyson96}
  Dell'Antonio, I.P. \& Tyson, J.A.\ 1996, ApJ 473, L17
\bibitem[1997]{DolagBartelmann}
Dolag, K. \& Bartelmann, M.\ 1997, MNRAS 291, 446
\bibitem[2000]{Fischer00}
  Fischer, P., McKay, T.A., Sheldon, E.\ et al.\ 2000, AJ 120, 1198
\bibitem[1996]{Griffiths96}
  Griffiths, R.E., Casertano, S., Im, M. \& Ratnatunga, K.U.\ 1996,
  MNRAS 282, 1159
\bibitem[2002]{GuSel02}
 Guzik, J. \& Seljak, U.\ 2002, MNRAS 335, 311
\bibitem[2002c]{hoek604}
  Hoekstra, H., van Waerbeke, L., Gladders, M.D., Mellier, Y. \& Yee,
  H.K.C.\ 2002, ApJ 577, 604
\bibitem[2001]{Hoek01}
  Hoekstra, H., Yee, H.K.C. \& Gladders, M.D.\  2001, ApJ 558, L11
\bibitem[1998]{Hudson98}
  Hudson, M.J., Gwyn, S.D.J., Dahle, H. \& Kaiser, N.\ 1998,
  ApJ 503, 531
\bibitem[1997]{jaselj}
  Jain, B. \& Seljak, U. 1997, ApJ, 484, 560
\bibitem[2003]{Jarvisetal1}
  Jarvis, M., Bernstein, G.M., Fischer, P.\ et al.\ 2003, AJ 125. 1014
\bibitem[2003]{Jarvisetal}
  Jarvis, M., Bernstein, G.M. \& Jain, B.\ 2004, MNRAS 352, 338
\bibitem[1992]{kai92}
  Kaiser, N. 1992, ApJ, 388, 272
\bibitem[2000]{KWL00}
  Kaiser, N., Wilson, G. \& Luppino, G. 2000, astro-ph/0003338
\bibitem[2004]{KH04}
  Kleinheinrich, M., Rix, H.-W., Erben, T., et al.\ 2004, A\&A
  submitted (also astro-ph/0404527)
\bibitem[2001]{Maoli01}
  Maoli, R., van Waerbeke, L., Mellier, Y.\ et al.\ 2001, A\&A 368, 766
\bibitem[2001]{McKay01}
  McKay, T.A., Sheldon, E.S., Racusin, J.\ et al.\ 2001, astro-ph/0108013
\bibitem[1999]{Me99}
  Mellier, Y. 1999, ARA\&A, 37, 127
\bibitem[2003]{Menardetal}
  M\'{e}nard, B., Bartelmann, M. \& Mellier, Y.\ 2003, A\&A 409, 411
\bibitem[1991]{MiEs}
  Miralda-Escud\'e, J. 1991, ApJ 380, 1
\bibitem[1980]{PeeblesLSS}
Peebles, P.J.E.\ 1980, {\it The large-scale structure of the
universe}, Princeton University Press
\bibitem[2002]{Penetal02}
  Pen, U.-L., Zhang, T., van Waerbeke, L., Mellier, Y., Zhang, P. \&
  Dubinski, J. 2003, ApJ 592, 664
\bibitem[2003a]{Refre03a}
  Refregier, A.\ 2003, ARA\&A 41, 645
\bibitem[2002]{RRG02}
  Refregier, A. Rhodes, J. \& Groth, E.J.\ 2002, ApJ 572, L131
\bibitem[1996]{S96}
  Schneider, P.\ 1996, MNRAS, 283, 837
\bibitem[1998]{S98}
  Schneider, P.\ 1998, ApJ 498, 43
\bibitem[2003]{S03}
  Schneider, P.\ 2003, A\&A 408, 829
\bibitem[2004]{SKL04}
  Schneider, P., Kilbinger, M. \& Lombardi, M.\ 2004, A\&A, submitted
  (also astro-ph/0308328)
\bibitem[2003]{SL03}
  Schneider, P. \& Lombardi, M.\ 2003, A\&A 397, 809
\bibitem[2002]{SvWM02}
  Schneider, P., van Waerbeke, L. \& Mellier, Y.\ 2002, A\&A 389, 729
\bibitem[1998]{SvWJK}
  Schneider, P., van Waerbeke, L., Jain, B., Kruse, G., 1998, MNRAS, 296,
  873
\bibitem[2004]{Selj04}
  Seljak, U., Makarov, A., Mandelbaum, R.\ et al.\ 2004, astro-ph/0406594
\bibitem[2004]{Sheldon04}
  Sheldon, E.S., Johnston, D.E.,  Frieman, J.A.\ et al.\ 2004,
  AJ 127, 2544
\bibitem{takajain03}
  Takada, M. \& Jain, B.\ 2003a, ApJ 583, L49
\bibitem{takajain03b}
  Takada, M. \& Jain, B.\ 2003b, MNRAS 344, 857
\bibitem[1984]{Tys84}
  Tyson, J.A., Valdes, F., Jarvis, J.F. \& Mills~Jr., A.P.\ 1984,
  ApJ 281, L59
\bibitem[1998]{Ludo98}
  Van Waerbeke, L.\ 1998, A\&A 334, 1
\bibitem[1999]{vWBM99}
  van Waerbeke, L., Bernardeau, F. \& Mellier, Y. 1999, A\&A, 243, 15
\bibitem[2000]{vWetal00}
  Van Waerbeke, L., Mellier, Y., Erben, T.\ et al. 2000, A\&A, 358, 30
\bibitem[2001]{vWetal01}
  Van Waerbeke, L., Mellier, Y., Radovich, M.\ et al. 2001, A\&A, 374, 757
\bibitem[2003]{vW-M-Review03}
van Waerbeke, L. \& Mellier, Y.\ 2003, astro-ph/0305089
\bibitem[2002]{vW02etal}
van Waerbeke, L., Mellier, Y., Pello, R., et al.\ 2002, A\&A 393, 369
\bibitem[2000]{Wittm00}
  Wittman, D.M., Tyson, J.A., Kirkman, D., Dell'Antonio, I. \&
  Bernstein, G. 2000, Nat, 405, 143
\bibitem[2003]{ZalSco03}
  Zaldarriaga, M. \& Scoccimarro, R.\ 2003, ApJ 584, 559

\end{thebibliography}
\end{document}